\documentclass[twoside]{article}
\usepackage{fleqn,epsfig,espcrc2}

\usepackage{graphicx}
\usepackage[figuresright]{rotating}

\newcommand{\AmS}{{\protect\the\textfont2
  A\kern-.1667em\lower.5ex\hbox{M}\kern-.125emS}}
\def\beq{\begin{equation}}
\def\eeq{\end{equation}}
\def\bea{\begin{eqnarray}}
\def\eea{\end{eqnarray}}
\def\bq{\begin{quote}}
\def\eq{\end{quote}}
\def\gappeq{\mathrel{\rlap {\raise.5ex\hbox{$>$}}
{\lower.5ex\hbox{$\sim$}}}}

\def\lappeq{\mathrel{\rlap{\raise.5ex\hbox{$<$}}
{\lower.5ex\hbox{$\sim$}}}}

\def\Toprel#1\over#2{\mathrel{\mathop{#2}\limits^{#1}}}

\def\beaa{\begin{eqnarray*}}
\def\eeaa{\end{eqnarray*}}

\title{Astroparticle Aspects of Supersymmetry}

\author{John Ellis\address{Theoretical Physics Division, CERN, CH-1211
Geneva 23, Switzerland}}
       
\begin{document}

\begin{abstract}
After recalling the motivations for expecting supersymmetry to appear at
energies $\lappeq 1$~TeV, the reasons why the lightest supersymmetric
particle is an ideal candidate for cold dark matter are reviewed from a
historical perspective. Recent calculations of the relic density including
coannihilations and rapid annihilations through direct-channel Higgs boson
poles are presented. The experimental constraints from LEP and elsewhere
on supersymmetric dark matter are reviewed, and the prospects for its
indirect or direct detection are mentioned. The potential implications of
a Higgs boson weighing about 115~GeV and the recent measurement of the
anomalous magnetic moment of the muon are summarized.
\vspace{1pc}
\end{abstract}

\maketitle

\begin{center}
CERN-TH/2001-091 \qquad \qquad hep-ph/0103288
\end{center}

\section{Why Supersymmetry?}

As we have heard at this meeting, theorists have been attracted to
supersymmetry for many different reasons: because it is there, because it
is beautiful, to unify theories of matter and interactions, etc. However,
these profound theoretical arguments have not been able to fix the energy
scale where supersymmetry should appear, which could, {\it a priori}, be
as large as the Planck scale $\sim 10^{19}$~GeV. On the other hand, an
argument for fixing the masses of the supersymmetric partners of the
Standard Model particles to be around 1~TeV is provided by the chance
offered by supersymmetry of making a large hierarchy of mass scales more
natural~\cite{hierarchy}.

By now, there are several circumstantial experimental hints in favour of
supersymmetry at the TeV scale. One is the agreement of the gauge coupling
strengths measured at LEP and elsewhere with supersymmetric grand
unification, if the spartners of the Standard Model particles weigh $\sim$
1 TeV~\cite{susyGUT}. Another indication is provided by precision
electroweak data, that favour a light Higgs boson~\cite{lightH}, as
predicted in the MSSM~\cite{EFZ}~\footnote{Later in this talk, the
possibility that direct Higgs searches may be providing more than just an
indication is also discussed~\cite{EGNO}.}. Yet another motivation for new
physics at the TeV scale may have been provided recently by the anomalous
magnetic moment of the muon~\cite{BNL}, and supersymmetry is quite
capable~\cite{ENO} of explaining the 2.6-$\sigma$ discrepancy between
theory and the recent BNL E821 experiment~\cite{BNL}.

In this talk, however, I should like to emphasize a third, astrophysical,
motivation for TeV-scale supersymmetry, namely the cold dark matter for
which the lightest supersymmetric particle is an ideal candidate. Recall
that conventional (baryonic) matter constitutes $\lappeq$ 5 \% of the
critical density, according to analyses of Big-Bang
nucleosynthesis~\cite{BBN} and the cosmic microwave background (CMB) 
radiation~\cite{CMBR}.  There are both astrophysical and
particle-experimental reasons to believe that massive neutrinos could
provide at most a similar percentage of the critical density~\cite{nu}. 
However, analyses of the CMB and large-scale structure data suggest that
the total amount of matter in the Universe corresponds to about 30~\% of
the critical density, far more than could be provided by either baryons or
neutrinos~\cite{CMBR}. Theories of structure formation suggest that most
of the matter density should be in the form of cold dark matter~\cite{SF},
such as massive weakly-interacting particles.  The remaining 65 \% of the
critical density, as required in inflationary cosmology, indicated by data
on high-redshift supernovae~\cite{hiz} and supported by the CMB
data~\cite{CMBR}, is presumably provided by some form of vacuum energy,
either varying with time or a true cosmological constant.

Turning to particle candidates for the cold dark matter, massive
weakly-interacting particles that were originally in thermal
equilibrium could have interesting cosmological densities if they had
either (a) masses in the eV region, such as neutrinos, (b) masses of a few
GeV, now excluded by LEP for particles with neutrino-like couplings to the
$Z^0$~\cite{ERS}, or (c) masses in the range from $m_Z/2$ up to about 1
TeV. It is
the latter possibility that may be realized with the lightest
supersymmetric particle (LSP). 

In this talk, I first review how the realization grew that the LSP could
be an excellent candidate for cold dark matter. Then I review the
constraints on supersymmetric dark matter coming from accelerator
experiments, principally those from LEP. Then I mention the principal
strategies for searching for supersymmetric dark matter~\cite{sarches},
including
indirect searches for supersymmetric relic annihilations in the galactic
halo or core, or inside the Sun or Earth, and direct searches for the
scattering of dark matter particles on nuclei in the laboratory. Since one
of the tightest constraints on the mass of the LSP is provided by the
Higgs search at LEP, which has recently been the subject of much interest,
I include a discussion of the status of Higgs searches at LEP. Also, I
cannot resist adding some comments on the possible supersymmetric
interpretation~\cite{ENO} of the recent BNL E821 measurement of the muon
anomalous magnetic moment~\cite{BNL}. 

\section{The Lightest Supersymmetric Particle}

In many supersymmetric models, this is expected to be stable~\cite{Fayet},
and hence
likely to be present today in the Universe as a cosmological relic from
the Big Bang. It is stable because of a multiplicatively-conserved quantum
number called $R$ parity~\cite{Fayet}, which takes the value +1 for all
conventional
particles and -1 for all sparticles. Its conservation is linked to those
of baryon and lepton numbers: 
\beq
R = (-1)^{3B+L+2S}
\label{one}
\eeq
where $S$ is the spin. It is certainly possible to violate $R$ by
violating $L$ either spontaneously or explicitly, but these options are
limited by laboratory~\cite{Rviol} and cosmological
constraints~\cite{CDEO}, and we discard them
for the rest of this talk. 

The conservation of $R$ parity has the following three important consequences.
\\
1) Sparticles are always produced in pairs, e.g., $pp \rightarrow \tilde q 
\tilde q + X$ or $e^+e^-\rightarrow \tilde\mu^+\tilde\mu^-$.\\
2) Heavier sparticles decay into lighter ones, e.g., $\tilde q\rightarrow 
q\tilde q$ or $\tilde\mu\rightarrow\mu\tilde\gamma$.\\
3) The lightest supersymmetric particle is stable, because it has no legal 
decay mode.\\
It is the latter property that makes the LSP an ideal particle candidate
for dark matter.

Although Fayet~\cite{Fayet} had appreciated previously the significance of
$R$ parity, and he and Farrar~\cite{FF} had discussed the phenomenology in
laboratory experiments of the LSP, the first discussion of supersymmetric
cold dark matter was by Goldberg~\cite{Goldberg}, as far as I am aware. He
discussed the case where the LSP is a pure photino $\tilde \gamma$, and
linked its possible mass to those of the squarks. The next two papers
known to me were~\cite{EHNS} and~\cite{Krauss}. We saw the Goldberg
paper~\cite{Goldberg} while we were thinking how to search for
supersymmetry at the CERN ${\bar p} p$ collider, and were impressed by his
argument that cosmology imposed important constraints. However, we also
realized that the pure photino limit he considered was insufficient, since
this state must mix with the spartners of the neutral Higgs bosons and the
$Z^0$, in any realistic model.  We added some consideration of 
this mixing in our collider
paper~\cite{EHNS}, and resolved to study in more detail the full case of
mixing between the ${\tilde \gamma}, {\tilde H}$ and ${\tilde Z^0}$ in a
generic neutralino $\chi$. The resulting paper~\cite{EHNOS} came out a few
months later. 

In addition to calculations of the neutralino relic density, we also
presented in~\cite{EHNOS} some general arguments on the likely nature of
the LSP, in
particular why it should interact only weakly.
If the supersymmetric relic had either electric charge or strong interactions,
it
would have condensed into ordinary matter, and shown up as an anomalous heavy
isotope. These have not been seen, and experiments
impose~\cite{nocharged}
\beq
{n(relic)\over n(p)} \lappeq 10^{-15} ~~{\rm to}~~ 10^{-30}
\label{two}
\eeq
for 1 GeV $\lappeq m_{relic} \lappeq$ 10 TeV, far below the expected
supersymmetric relic abundance. We conclude that the supersymmetric relic is
surely electrically neutral and weakly-interacting, e.g., it cannot be a
gluino~\cite{EHNOS}. The latter possibility is resuscitated occasionally,
but we are convinced that it is excluded by this anomalous-isotope
argument.

Looking through the sparticle data book, the possible cold dark matter
scandidates are the sneutrinos ${\tilde \nu}$ of spin 0, the lightest
neutralino $\chi$ of spin 1/2, and the gravitino ${\tilde G}$ of spin 3/2.
Sneutrinos are excluded by a combination of LEP and direct searches.  The
gravitino is the LSP in gauge-mediated models, in particular, but would
only provide hot (or, at best, warm) dark matter, rather than cold, so we
do not discuss it further here.

The lightest neutralino $\chi$ is therefore the favoured dark matter
candidate in many supersymmetric models. At the tree level, it is
characterized by 3 parameters:  the unmixed gaugino mass, $m_{1/2}$,
assumed to be universal, the Higgs mixing parameter, $\mu$, and the ratio
of Higgs v.e.v.'s, $\tan\beta$~\cite{EHNOS}. The neutralino composition
simplifies in the limit $m_{1/2}\rightarrow 0$, where it becomes an almost
pure photino ${\tilde \gamma}$~\cite{Goldberg}, and in the limit
$\mu\rightarrow 0$, where it becomes an almost pure Higgsino ${\tilde
H}$~\cite{EHNS}. However, both of these limits are excluded by LEP, which
enforces $m_\chi \gappeq$ 100 GeV~\cite{EFGO,EGNO,EFGOSi} and non-trivial
mixing, as we
discuss in more detail below. 

One of the most exciting features of neutralino dark matter is that there
are generic domains of parameter space where an `interesting' cosmological
relic density: $0.1 \lappeq \Omega_\chi h^2 \lappeq 0.3$ is possible for
some suitable choice of the other supersymmetric model
parameters~\cite{EHNOS,EFGO}. This point will be discussed in more
detail below, after further discussion of the experimental and
cosmological constraints on supersymmetric dark matter. 

\section{Experimental and Cosmological Constraints}

\begin{figure}[htbp]
\begin{center}
\mbox{\epsfig{file=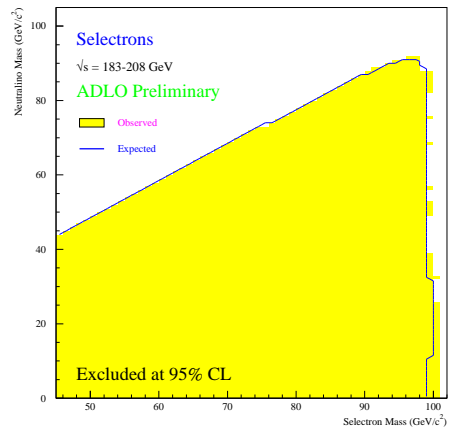,height=6.5cm}}
\end{center}
\caption[.]{\label{fig:selectron}\it
LEP constraint on the selectron, in the
$m_{\tilde e}, m_\chi$ plane~\cite{selectron}.
}
\end{figure}

\begin{figure}[htbp]
\begin{center}
\mbox{\epsfig{file=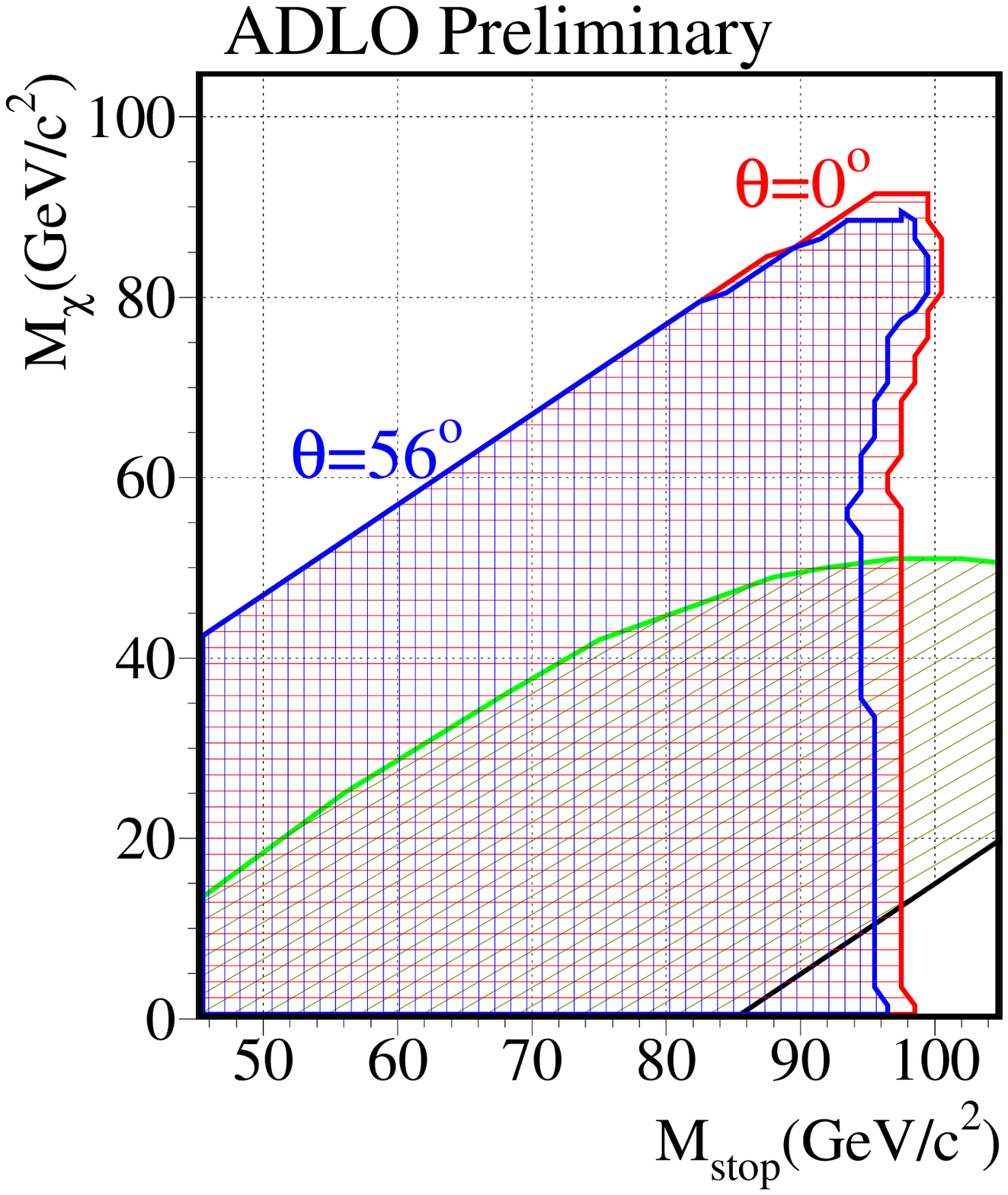,height=6.5cm}}
\end{center}
\vskip-2.0cm
\caption[.]{\label{fig:stop}\it
LEP constraints on the lighter stop ${\tilde t_1}$, in the
$m_{\tilde t_1}, m_\chi$ plane, for various mixing scenarios~\cite{stop}.
}
\end{figure}

The most direct limits on supersymmetric dark matter come from
direct searches for other supersymmetric particles. For example,
LEP has established that {\it charginos} $\chi^\pm$ weigh $\ge
103.5$~GeV~\cite{chargino},
and has also established important limits by searching for the
associated production of neutralinos: $ e^+e^-\rightarrow \chi +
\chi^\prime$. LEP has also established a lower limit of 100 GeV on
the {\it selectron} mass, as seen in
Fig.~\ref{fig:selectron}~\cite{selectron}. However, one of the most
stringent sparticle
limits comes indirectly from the {\it Higgs} search~\cite{Higgs},
since~\cite{EFGO} the
Higgs mass is sensitive to sparticle masses via radiative
corrections~\cite{EFZ}: 
\beq
\delta m^2_h \; \propto \; {m^4_t\over m^2_W}~\ln~\left({m^2_{\tilde
t}\over
m^2_t}\right)
\label{three}
\eeq
Because of the sensitivity to $m_{\tilde t}$ in (\ref{two}), experimental
constraints on {\it squarks}, in particular the lighter {\it
stop}: $m_{\tilde t_1} \ge 95$~GeV, as seen in
Fig.~\ref{fig:stop}~\cite{stop}, also
impact the neutralino limits.  

\begin{figure}[htbp]
\begin{center}
\mbox{\epsfig{file=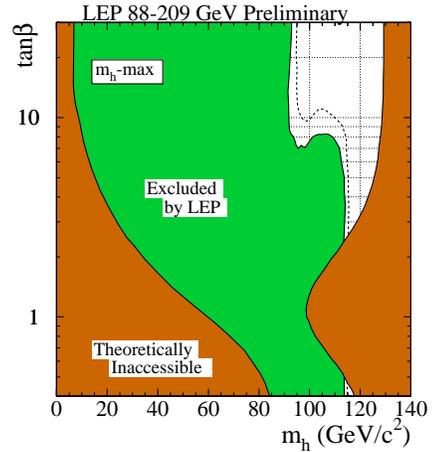,height=6.5cm}}
\end{center}
\caption[.]{\label{fig:Higgs}\it
LEP constraints on MSSM Higgs bosons, in the
$m_h, \tan\beta$ plane, assuming a maximal mixing scenario and
neglecting CP violation~\cite{Higgs}.
}
\end{figure}

\begin{figure}
\vspace*{-0.25in}
\begin{minipage}{8in}
\epsfig{file=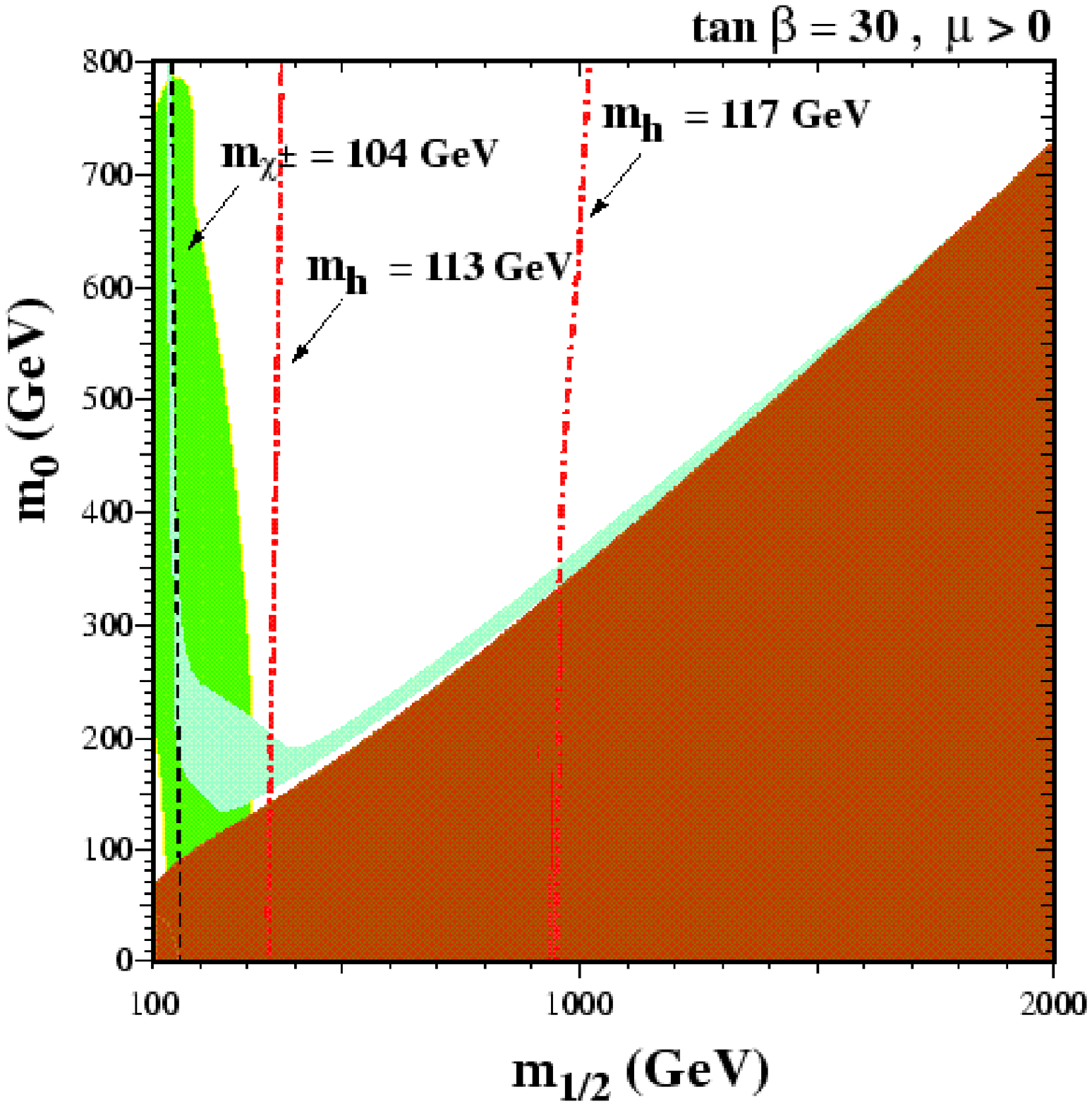,height=2.75in}
\hfill
\end{minipage}
\begin{minipage}{8in}
\epsfig{file=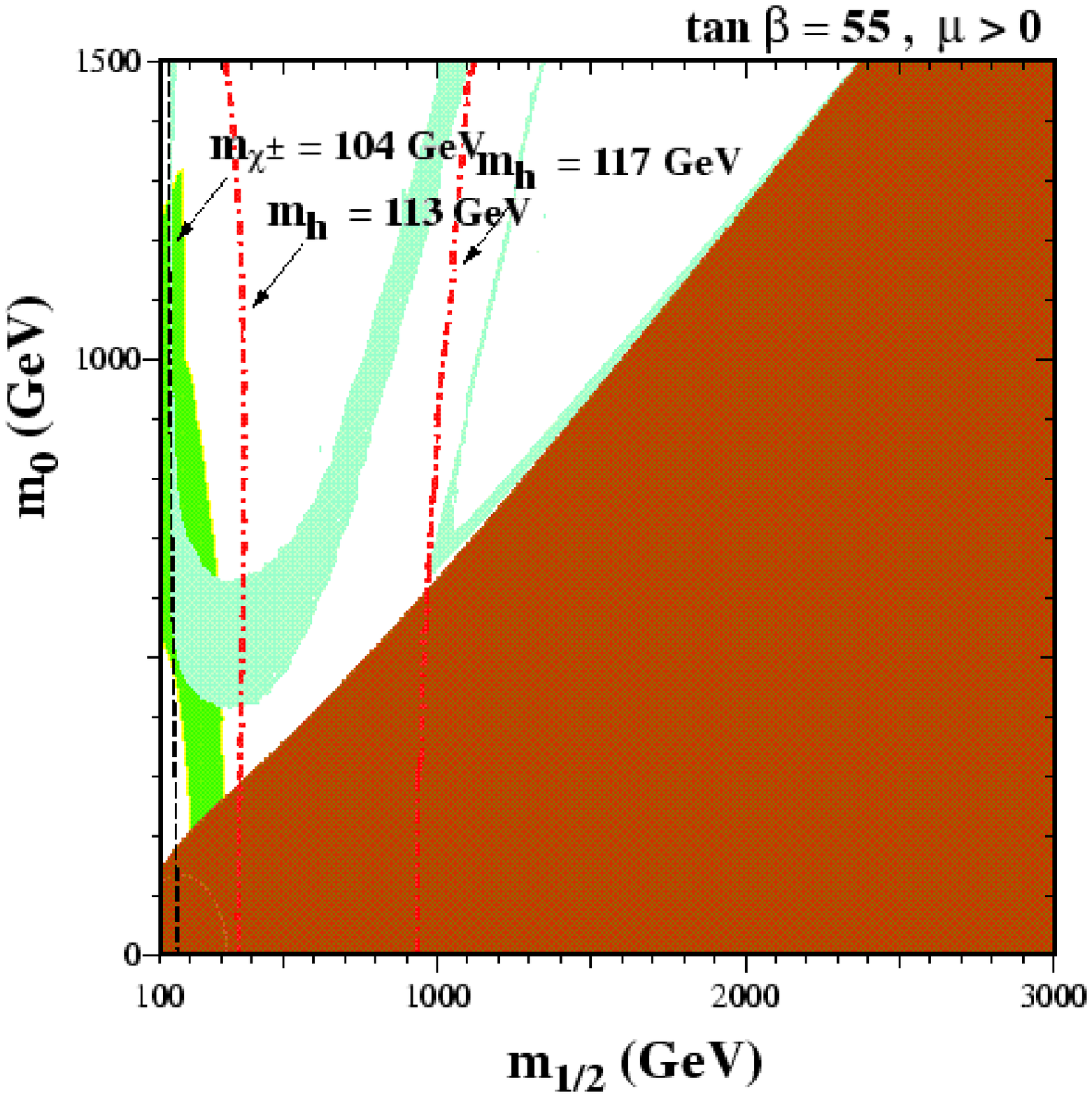,height=2.75in}
\hfill
\end{minipage}
\caption[.]{\label{fig:EFGOSiplus}\it
Domains of the $m_0, m_{1/2}$ plane in the CMSSM for $\mu > 0$,
and $\tan \beta =$ (a) 30, (b) 55, showing the impacts of various
experimental and cosmological
constraints~\cite{EFGOSi}.
Vertical broken lines indicate Higgs mass contours, assuming universality
of the
soft supersymmetry-breaking scalar mass $m_0$ and gaugino mass $m_{1/2}$.
Also shown in panel (a) is the impact of chargino searches at LEP. The
dark shaded region
is excluded because the lightest supersymmetric particle is the charged
${\tilde \tau}_1$, and
the medium shaded region is excluded by the measured $b \rightarrow s
\gamma$ rate. The
light shaded region is that preferred by cosmology, namely
$0.1 < \Omega h^2 < 0.3$.
}
\end{figure}

\begin{figure}
\vspace*{-0.15in}
\begin{minipage}{8in}
\epsfig{file=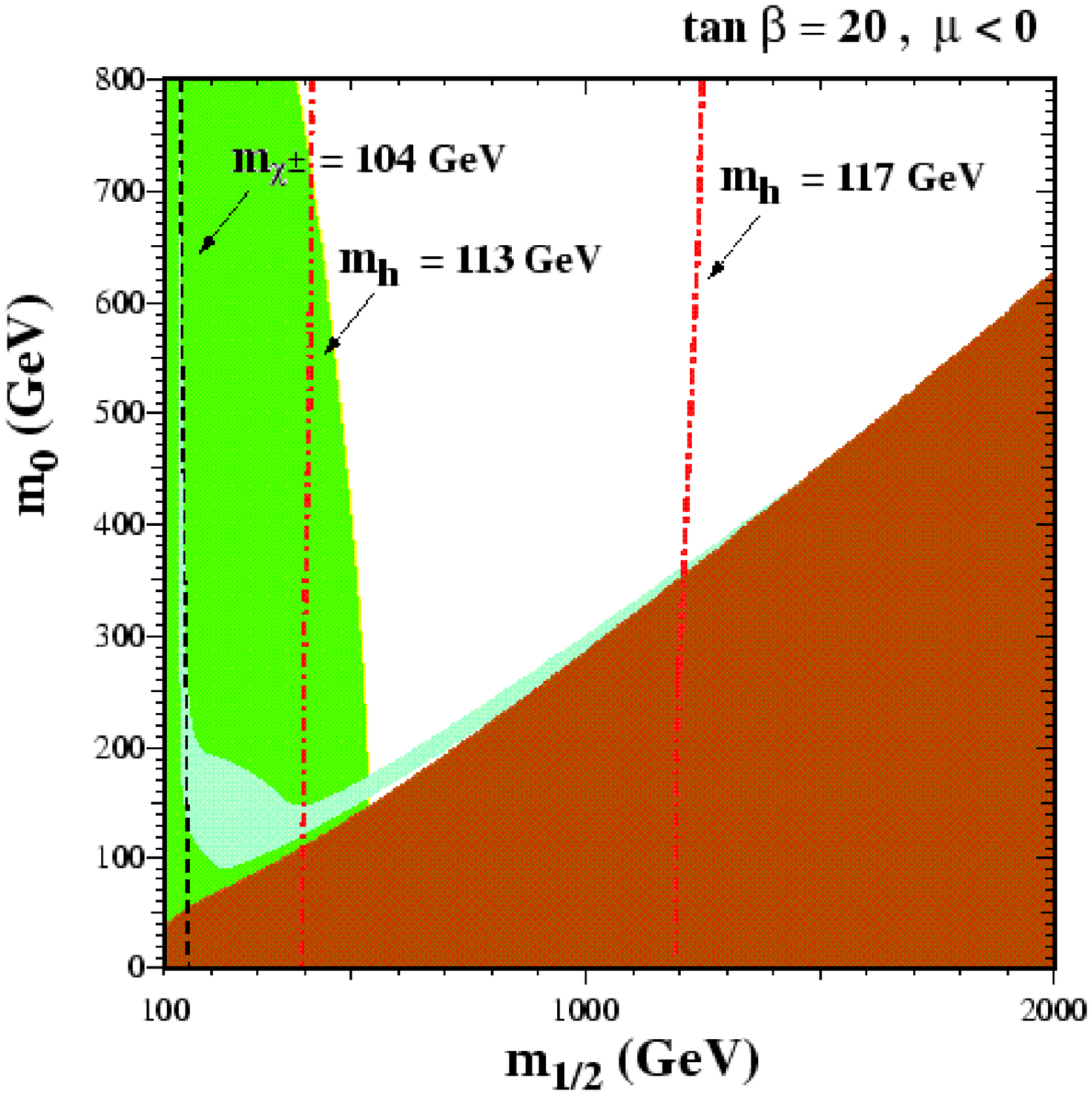,height=2.75in}
\hfill
\end{minipage}
\begin{minipage}{8in}
\epsfig{file=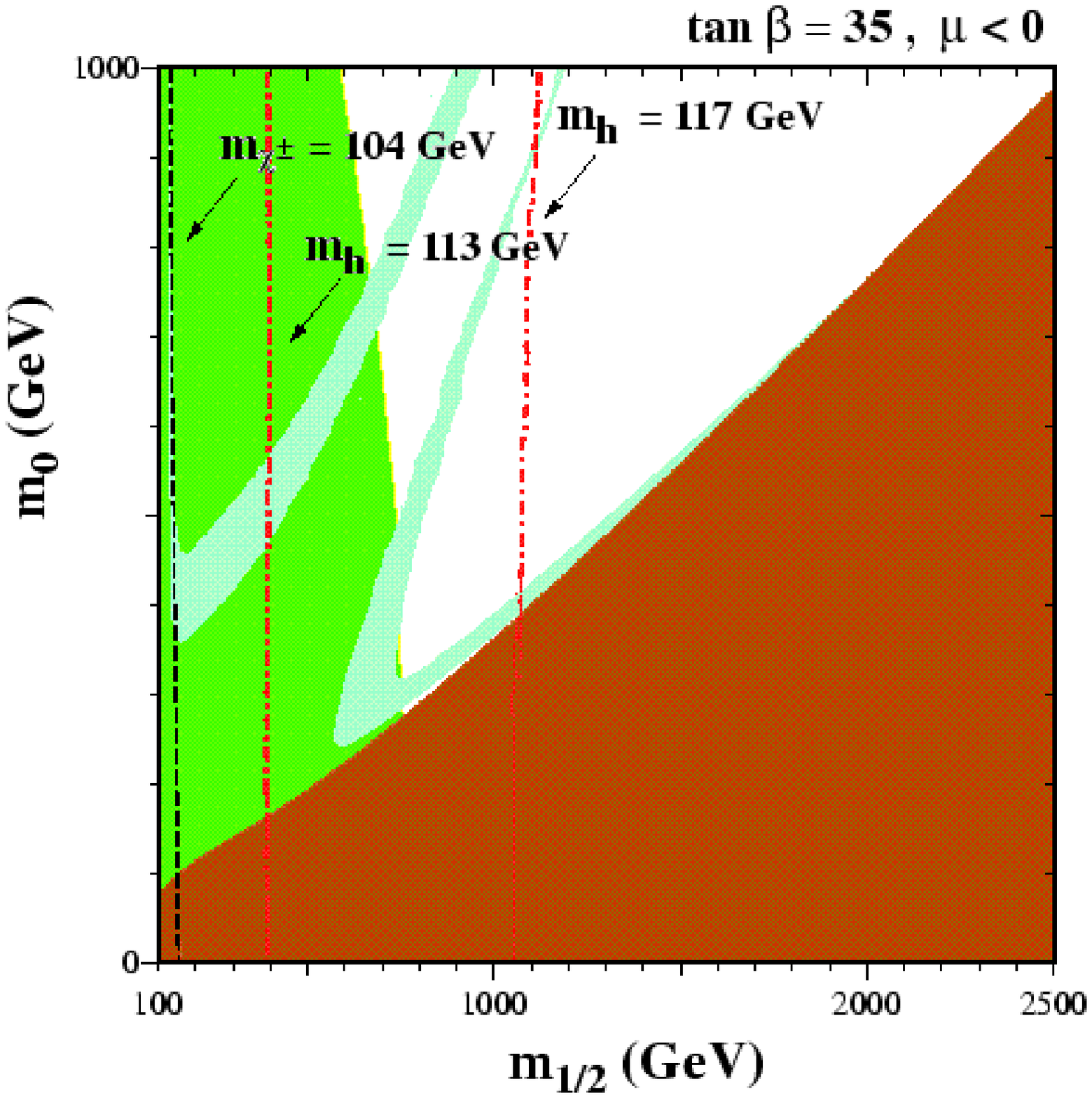,height=2.75in}
\hfill
\end{minipage}
\caption[.]{\label{fig:EFGOSiminus}\it
Domains of the $m_0, m_{1/2}$ plane in the CMSSM for $\mu < 0$
and $\tan \beta =$ (a) 20, (b) 35, showing the impacts of various
experimental and cosmological
constraints~\cite{EFGOSi}. Notations are the same as in
Fig.~\ref{fig:EFGOSiplus}.
}
\end{figure}

Relating the Higgs and the direct experimental
limits on different types of sparticles requires making some hypotheses on
the space of supersymmetric parameters.  The minimal supersymmetric
extension of the Standard Model (MSSM) is generally assumed, and sample
MSSM Higgs limits are shown in Fig.~\ref{fig:Higgs}. These are calculated
in a maximal mixing scenario, in which the $ZZh$ coupling may be
suppressed at large $\tan\beta$, relaxing the lower limit on $m_h$ from
its Standard Model value. In this talk I further assume the constrained
MSSM (CMSSM), in which the soft supersymmetry-breaking scalar and
fermionic mass parameters $m_{1/2}$ and $m_0$ are universal at some input 
GUT scale, as are the
soft trilinear supersymmetry-breaking parameters $A_0$. In this case, the
$ZZh$ coupling is not suppressed at large $\tan\beta$, and the LEP lower
limit on the lightest CMSSM Higgs boson is essentially identical to that
in the Standard Model. Representative examples of the sensitivity of
neutralino limits to Higgs and other limits are shown in
Figs.~\ref{fig:EFGOSiplus} and \ref{fig:EFGOSiminus}~\cite{EFGOSi}, for
$\mu > 0$ and $\mu < 0$, respectively.

We have included in Figs.~\ref{fig:EFGOSiplus} and
\ref{fig:EFGOSiminus}~\cite{EFGOSi} the constraint imposed by the
measurement of $b \rightarrow s \gamma$~\cite{bsgamma}, when compared with
the latest NLO
QCD calculations valid for large $\tan \beta$~\cite{NLObsg}, which is
shown as a
mdedium-shaded region. These plots also include the requirement that the
lighter stau, $\tilde\tau_1$, not be the LSP~\cite{EFGO}, which is
disallowed because the LSP cannot be charged. Fig.~\ref{fig:EFGOSiplus}(a)
also shows the impact of LEP chargino searches. If one is nervous about
the fate of our vacuum, one may also require that the effective potential
have no charge- and colour-breaking (CCB) minimum~\cite{CCB}. However,
this requirement may be regarded as optional: it is possible that our
present electroweak vacuum may be unstable but long-lived. Hence the CCB
constraint is not shown in Figs.~\ref{fig:EFGOSiplus} and
\ref{fig:EFGOSiminus}.

The light-shaded cosmological region in Figs.~\ref{fig:EFGOSiplus} and
\ref{fig:EFGOSiminus} has a `tail'
extending up to $m_{1/2} \sim$ 1400 GeV for $\tan \beta \lappeq 20$, as
seen in panels (a) of these figures,
which is due to efficient $\chi -
\tilde\ell$ coannihilations~\cite{EFOSi,othercoann}. These increase the
upper limit
on the LSP mass to
\beq
m_\chi \sim 600 {\rm GeV}.
\label{upperchi}
\eeq 
LHC appears able to find at
least some supersymmetric particles over all the comsologically-allowed
range for $\tan \beta \lappeq 20$~\cite{LHC}. However, for larger $\tan
\beta$, the
coannihilation region extends to larger $m_{1/2}$, as seen in panels
(b) of Figs.~\ref{fig:EFGOSiplus} and
\ref{fig:EFGOSiminus}~\footnote{In these panels, we have chosen the
largest values of $\tan \beta$ for which we consistent electroweak vacua
with our default choices of auxiliary parameters $m_t =
175$~GeV, $m_b(m_b)^{\overline MS}_{SM} = 4.25$~GeV and $A = 0$.}.
Hence, the `guarantee' of
supersymmetric discovery at the LHC does not extend to such larger values
of $\tan \beta$. Another feature visible for larger $\tan \beta$ is a
double `funnel' region of allowed LSP density extending out to large $m_0$
and $m_{1/2}$, which occurs because of rapid $\chi \chi \rightarrow A, H$
annihilation. This also extends beyond the region of `guaranteed'
discovery at the LHC~\footnote{At least within the CMSSM,
sparticle searches at Run II of the FNAL Tevatron collider will not be
able to probe much of the parameter space not already
excluded~\cite{Tevatron}.}. 

\begin{figure}[htb]
\begin{minipage}{8in}
\epsfig{figure=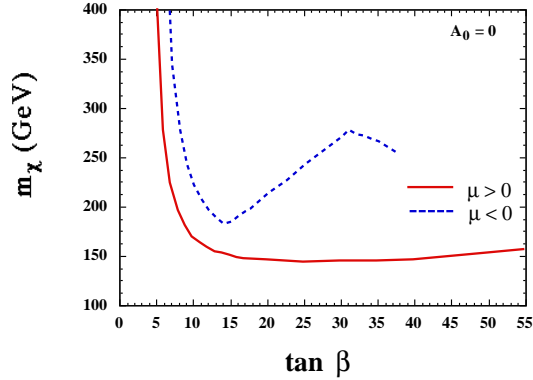,height=2.0in}
\end{minipage}
\caption[.]{\label{fig:lowerlimit}\it
Lower limits on the mass of the lightest supersymmetric particle in the
CMSSM, assuming $m_t =
175$~GeV, $m_b(m_b)^{\overline MS}_{SM} = 4.25$~GeV 
and $A_0 = 0$.
}
\end{figure}

The constraints
from LEP
and $b\rightarrow s \gamma$ impose a lower limit on $m_\chi$ within this
cosmological regionw. Its strength
depends
on $\tan\beta$ and theoretical assumptions, but we find in general
that~\cite{EFGOSi}
\beq
m_\chi \gappeq 100~{\rm GeV~~~and}~~~\tan\beta \gappeq 3
\label{four}
\eeq
with the precise numbers depending on the sign of $\mu$ and different
assumptions, e.g., on the values of $m_t, m_b$ and $A_0$. If the pole
mass $m_t =
175$~GeV, the running mass $m_b(m_b)^{\overline MS}_{SM} = 4.25$~GeV
and $A_0 = 0$, we find $m_\chi \gappeq 140 (180)$~GeV and $\tan\beta
\gappeq 5 (7)$ for $\mu > (<) 0$, as seen in
Fig.~\ref{fig:lowerlimit}~\footnote{Note, also,
that the bound (\ref{four}) could be
relaxed if the soft
supersymmetry-breaking scalar masses are not universal.}.

Within the region accessible to the LHC, one may ask whether it will find
supersymmetry before the searches for astrophysical dark
matter~\footnote{As we see later, the E821 measurement~\cite{BNL} offers
some hope
of restoring the `guarantee' that the LHC will find supersymmetry.}.

\section{Searches for Dark Matter}

One strategy is to look for the annihilations of relic particles in the
galactic halo~\cite{halo}, which may yield observable fluxes of (stable)
particles
such as $\bar p, \gamma$ and $e^+$. Measurements of the low-energy $\bar
p$ flux already rule out some supersymmetric models~\cite{pbars}.
On the other hand, $\gamma$ searches do not yet rule out any models,
unless the relic density in our galactic halo is strongly
clumped~\cite{clumpy}. There
have been some reports of an excess of cosmic-ray positrons, but some of
these have now been contradicted, and they cannot be taken as evidence for
supersymmetric dark matter.

\begin{figure}
\vspace*{-1in}
\begin{minipage}{8in}
\vskip -1in
\epsfig{file=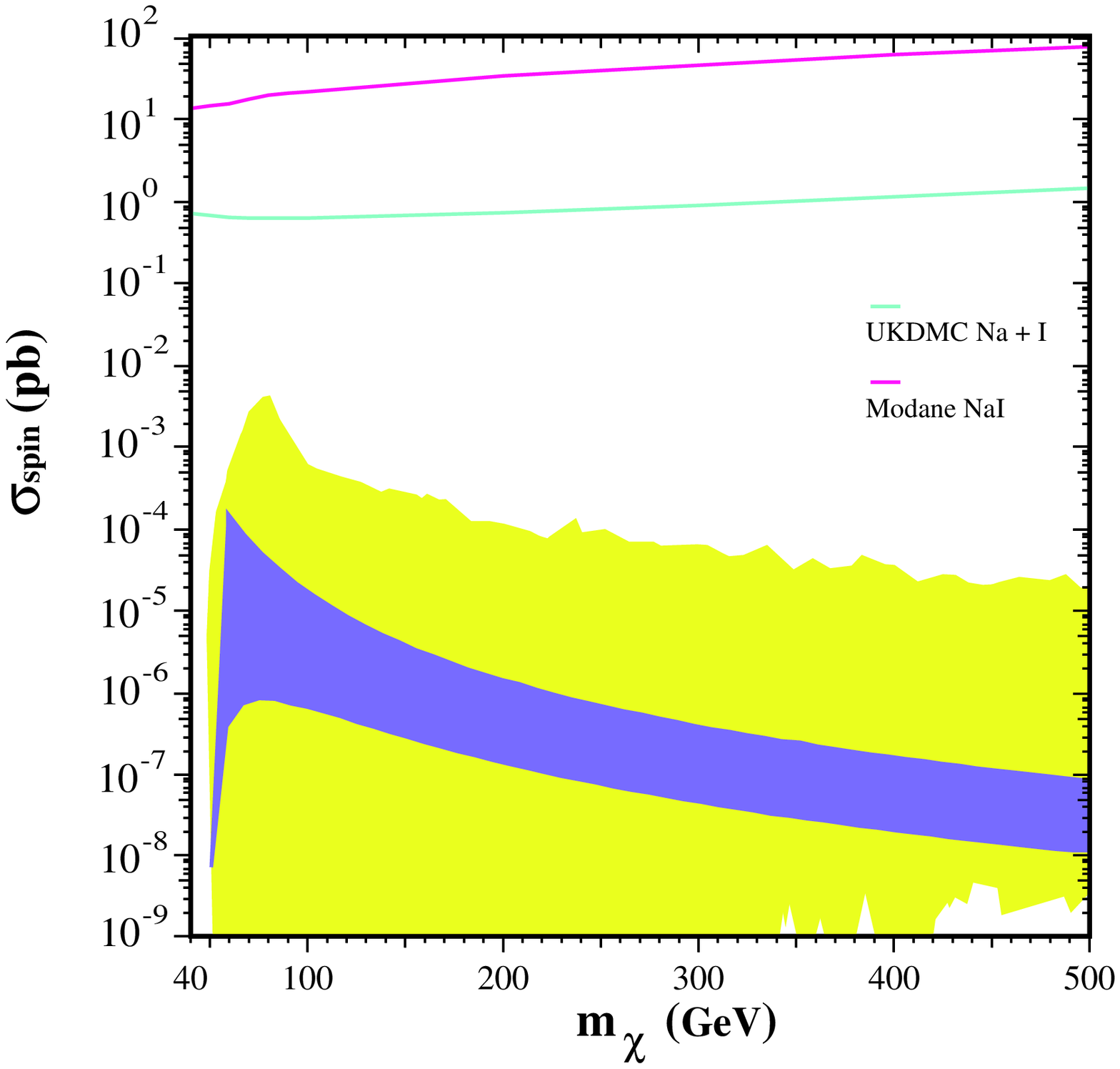,height=4in}
\hfill
\end{minipage}
\vspace*{-1in}
\begin{minipage}{8in}
\vskip -1in
\epsfig{file=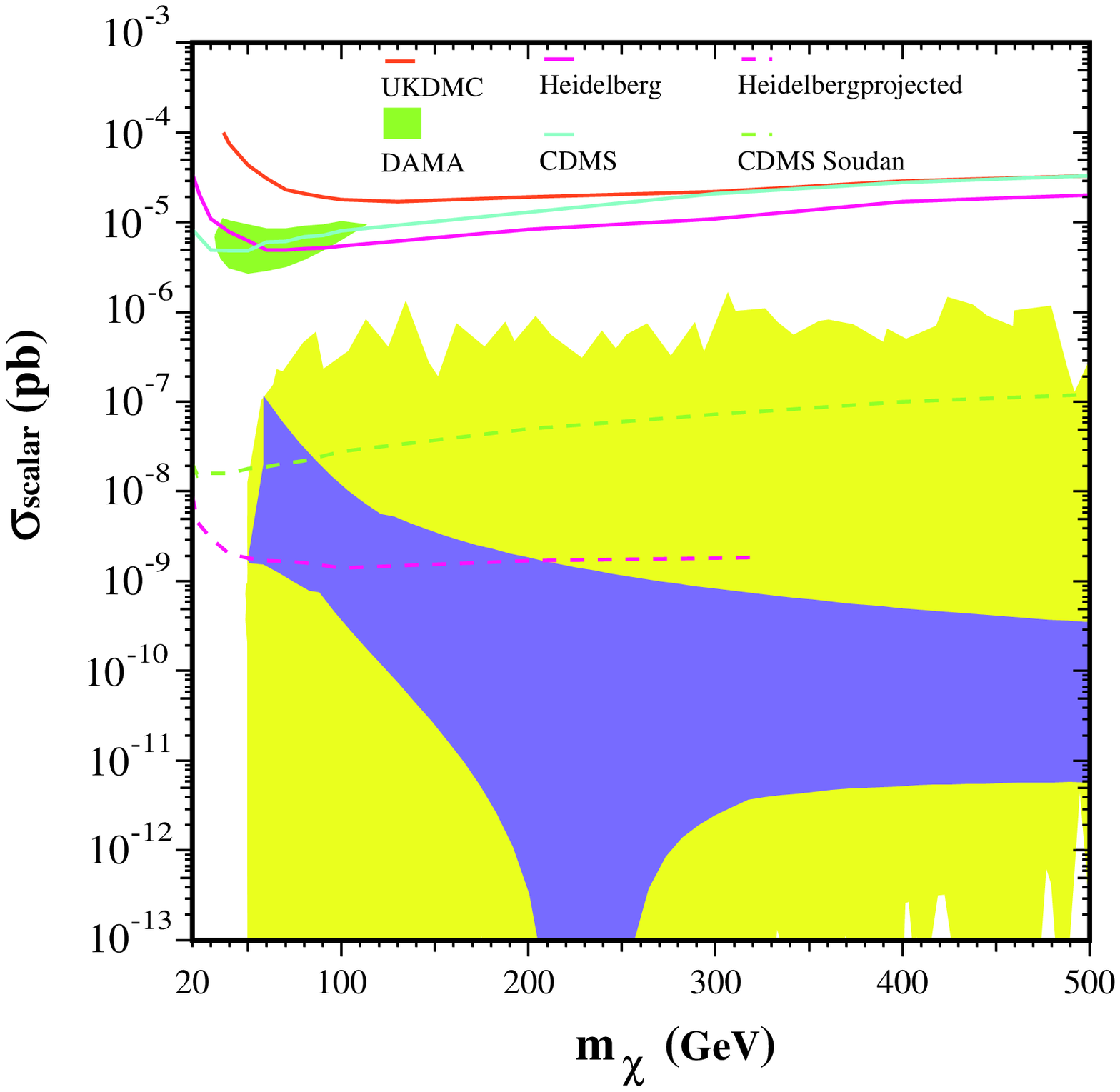,height=4in}
\hfill
\end{minipage}
\vskip 1.0cm
\caption[.]{\label{fig:EFO}\it
Compilations of ranges of the elastic cross sections found
in a sampling of supersymmetric models for $\tan \beta \le 10$, compared
with the sensitivity of
the DAMA~\cite{DAMA} and other experiments, for spin-independent 
scattering. The dark (light) shaded regions correspond to CMSSM
(MSSM) models with
(without) universal soft scalar supersymmetry-breaking
masses~\cite{EFO2}.}
\end{figure}

Another search strategy is to look for annihilations of relic particles
inside the Sun or Earth, following their capture after losing energy via
elastic scattering~\cite{Sun}. The observable products of these
annihilations are
high-energy neutrinos, which may either be seen directly in underground
detectors, or indirectly via $\nu$ collisions in rock that produce
detectable muons. Searches for neutrino-induced muons from the Sun and
Earth already exclude a significant number of supersymmetric models,
and there are prospects for improving the
current search sensitivities with a 1 km$^2$ (or 1 km$^3$)
detector~\cite{Gondolo}. 

The third favoured strategy is to search directly for elastic dark matter
scattering on nuclei in the laboratory~\cite{GW}. Each scattering event
would deposit typically an energy $E\sim m_\chi b^2/2 \sim$ tens of keV.
There are two important types of interaction: spin-dependent, which is
sensitive to the different quark contributions to the nucleon spin, and
spin-independent, which is sensitive to the quark contributions to the
nucleon mass. The former is likely to be more important for light nuclei,
the latter for heavy nuclei.

One experiment reports annual modulation of the rate of energy deposition
in their detector which may be interpreted as a possible signal for dark
matter scattering~\cite{DAMA}. We recently re-evaluated the rate of
elastic scattering in a constrained supersymmetric model, in which all the
soft scalar supersymmetry-breaking masses are assumed equal at the input
unification scale~\cite{EFO1}.  As seen in Fig.~\ref{fig:EFO}, we found an
elastic scattering cross-section below the region of experimental
interest, at least for $\tan \beta \le 10$. Subsequently, a
second
experiment published an upper limit that excluded most of the interesting
region~\cite{CDMS}. We have looked again at the elastic scattering rate,
this time relaxing the assumption of universal scalar masses~\cite{EFO2}.
Although the rates could be somewhat higher than in the universal case,
they still fell short of the region of experimental interest, at least for
$\tan \beta \le 10$~\cite{Bottino}~\footnote{We have recently extended
our analysis to larger $\tan \beta$ in the CMSSM~\cite{EFO3},
incorporating the latest
LEP experimental constraints, and find cross sections in a similar range
as before.}.

\section{What if ...?}

On September 5th, the ALEPH experiment at LEP reported an apparent excess
of Higgs boson candidates in the reaction $e^+e^-\rightarrow (H\rightarrow
\bar bb) + (Z \rightarrow \bar qq)$. Two of the ALEPH events
were rated as very good candidates that were quite unlikely to be due to
$e^+e^-\rightarrow Z + Z$, the most dangerous background reaction.
However, this background cannot be completely removed. During the period
up to Nov.~2nd, while LEP continued running, the LEP Higgs `signal' grew
approximately as would be expected if there was a Higgs boson weighing
about 115~GeV, with indications in other channels and in
other experiments~\cite{LEPHiggs}. However, a preliminary estimate of the
overall
significance of the reported signal only amounted to at most 2.9 standard
deviations~\cite{PIK}, not enough for LEP to claim discovery.
Nevertheless, one may speculate on the interpretation of any signal of a
Higgs boson weighing around 115 GeV~\cite{EGNO}.

The first implication is that the Standard Model must break down at some
scale $\lappeq 10^6$ GeV, with the appearance of some new bosonic
particles. This is because the Standard Model Higgs potential becomes
unstable at a scale $10^6$ GeV, because the radiative corrections
due to such a light Higgs boson are overwhelmed by those due to the
relatively heavy top quark. What sort of new physics may appear? 
Technicolour~\cite{TC} and other models with strongly-interacting Higgs
bosons are
excluded yet again~\cite{DE}, because they generically predict a heavier
Higgs
boson: 300 GeV $\lappeq m_H \lappeq$ 1 TeV. On the other hand, such a
light Higgs boson is completely consistent with supersymmetry, which
predicts that the lightest Higgs boson weighs $\lappeq$ 130
GeV~\cite{EFZ}. 

Moreover, it can be argued that the new bosonic physics must quack very
much like supersymmetry~\cite{ER}. It has to be quite finely tuned, if the
effective potential is not to blow up or become unbounded below. Such fine
tuning is natural in supersymmetry, but would be lost, e.g., if there were
no supersymmetric Higgsino partners of the Higgs boson. The discovery of a
Higgs boson weighing 115~GeV would therefore strengthen significantly the
circumstantial phenomenological case for supersymmetry. 

\begin{figure}
\begin{minipage}{8in}
\epsfig{file=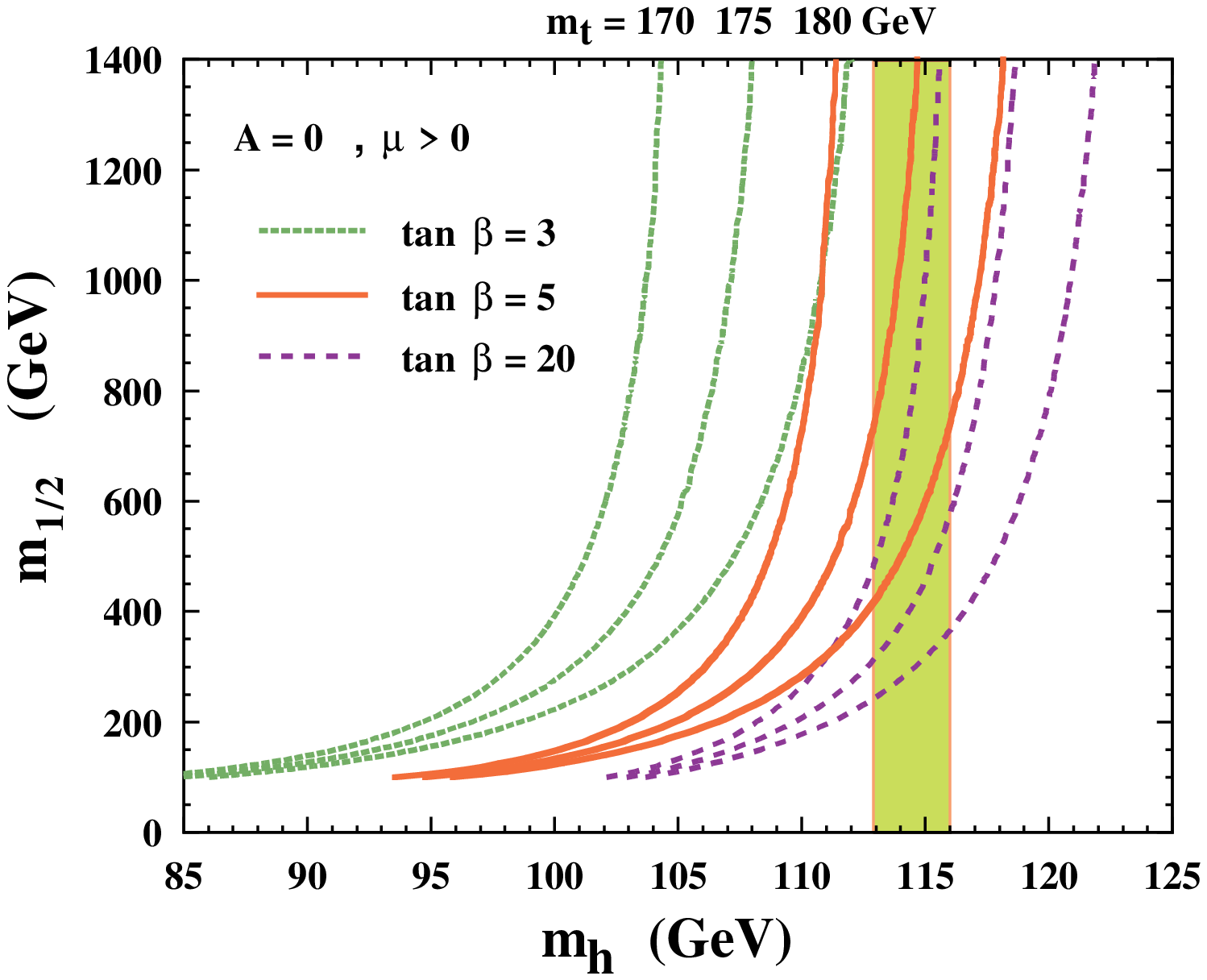,height=2in}
\hfill
\end{minipage}
\vspace*{-1in}
\begin{minipage}{8in}
\epsfig{file=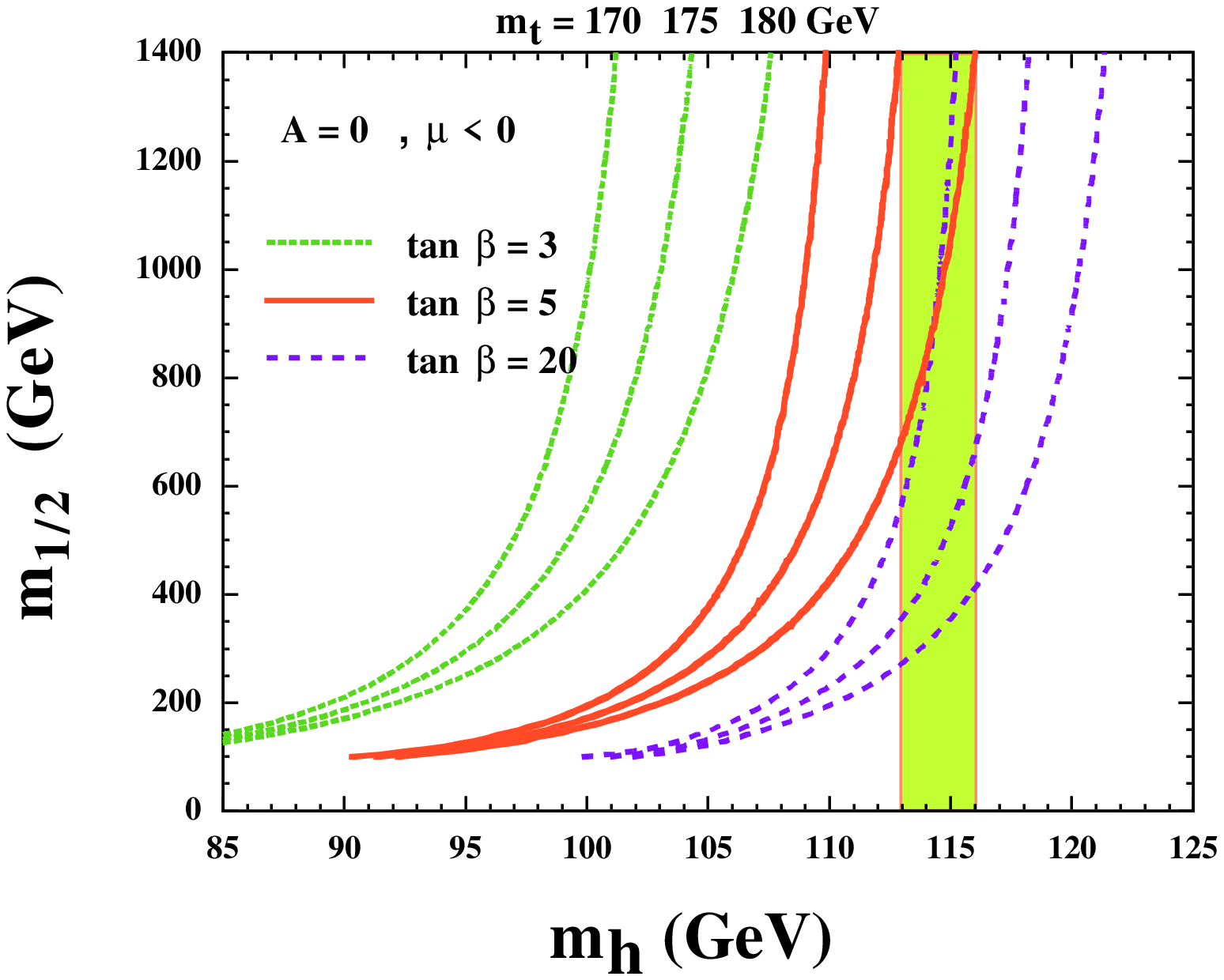,height=2in}
\hfill
\end{minipage}
\vskip 2.0cm
\caption[.]{\label{fig:EGNO}\it
The sensitivity of $m_h$ to $m_{1/2}$ in the CMSSM for (a) $\mu > 0$ and
(b) $\mu < 0$. The value $A_0 = 0$ is assumed for definiteness.
The dotted (green), solid (red) and dashed (blue) lines are for
$\tan \beta = 3, 5$ and $20$, each for $m_t = 170, 175$ and $180$~GeV
(from left to right). The lines are relatively unchanged as one varies
$\tan \beta \gappeq 10$, where they are also insensitive to the sign of
$\mu$.
The shaded vertical strip corresponds to $113~{\rm GeV} \le m_h   
\le 116~{\rm GeV}$.
}
\end{figure}

\begin{figure}
\begin{minipage}{8in}
\epsfig{file=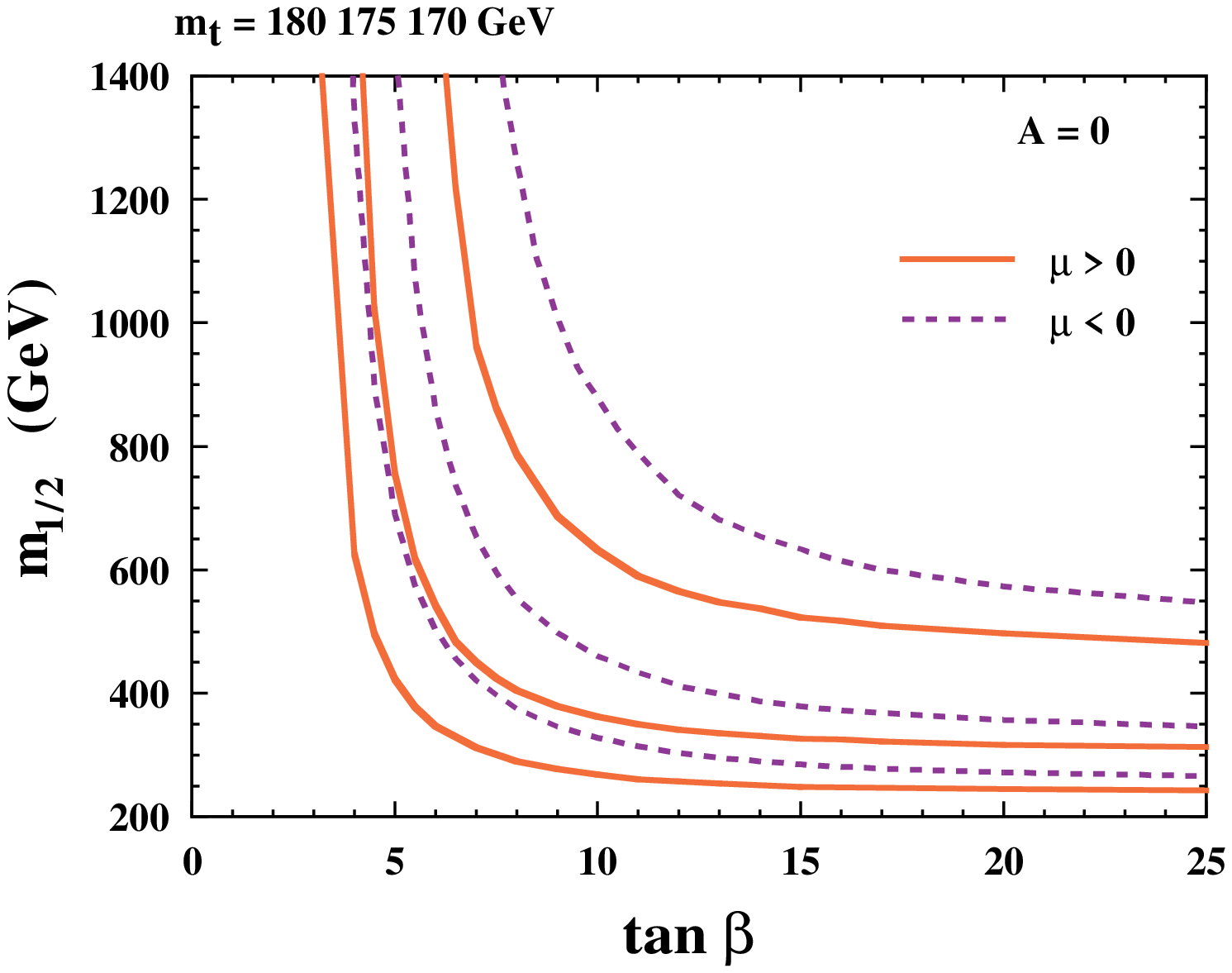,height=2in}
\hfill
\end{minipage}
\begin{minipage}{8in}
\epsfig{file=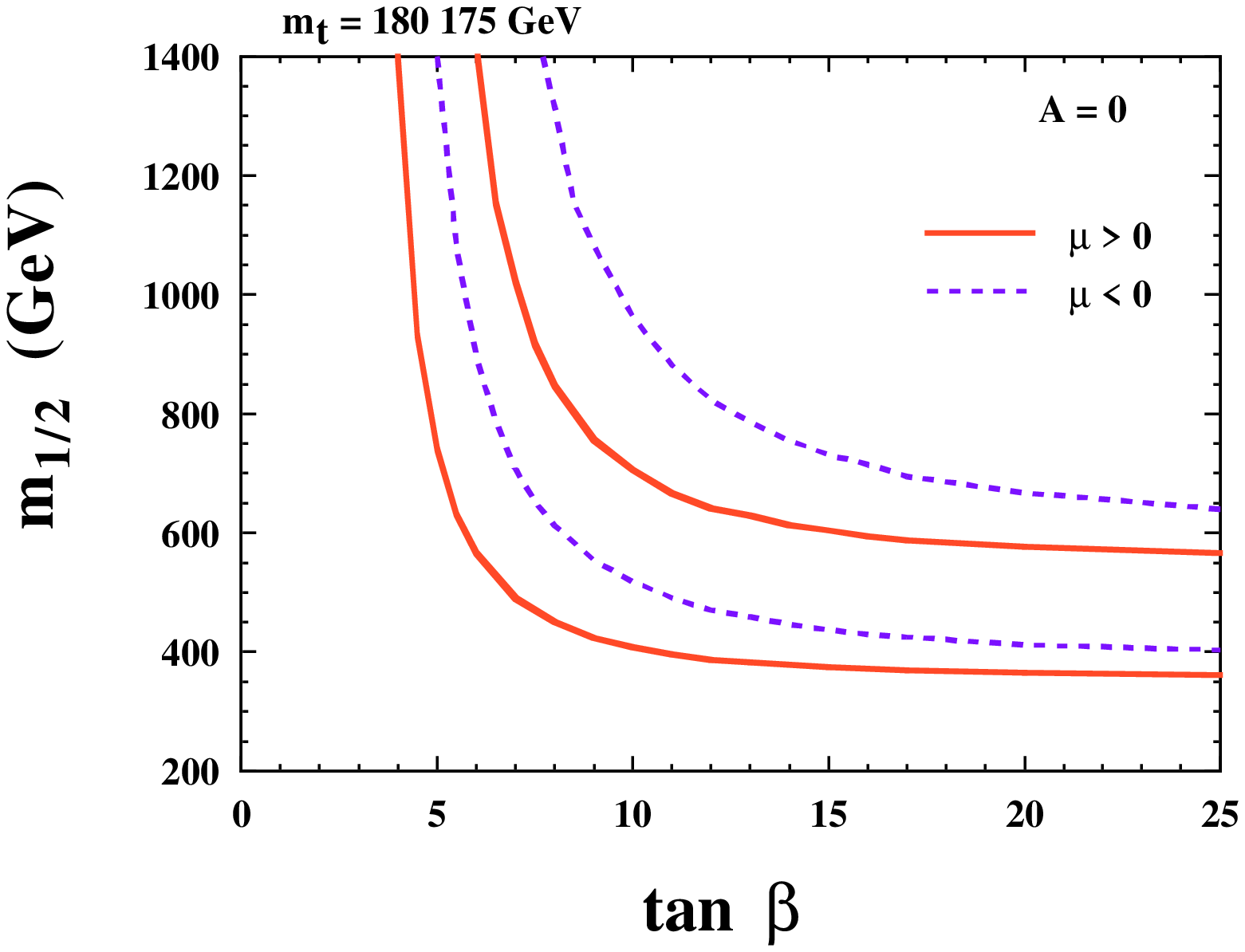,height=2in}
\hfill
\end{minipage}
\caption[.]{\label{fig:limit}\it
(a) The lower limit on $m_{1/2}$ required in the CMSSM to obtain $m_h \ge
113$~GeV for
$\mu > 0$ (solid, red lines) and $\mu < 0$ (dashed, blue lines), and $m_t
= 170, 175$ and $180$~GeV, and (b) the upper limit on $m_{1/2}$ required
to obtain $m_h \le 116$~GeV for both signs of $\mu$ and $m_t = 175$ and
$180$~GeV: if $m_t = 170$~GeV, $m_{1/2}$ may be as large as the
cosmological upper limit $\sim 1400$~GeV.  The corresponding values of the
lightest neutralino mass $m_\chi \simeq 0.4 \times m_{1/2}$. 
}
\end{figure}

In fact, as seen in Fig.~\ref{fig:EGNO}, a
measurement
of a Higgs mass around 115 GeV
could be used to estimate the value of $m_{1/2}$ required in the radiative
correction (\ref{three}). This is compatible with the range given earlier
(\ref{four}): if the present `signal' does not eventually
evaporate, the LSP mass may be quite close to the lower limit shown in
Fig.~\ref{fig:lowerlimit} as a function of $\tan \beta$. The principal
uncertainty in the estimate of $m_\chi$ is that due to uncertainties
in calculating $m_h$, in particular that due to the
range
$m_t = 175 \pm 5$ GeV, since $\delta m_h / \delta m_t = {\cal O}(1)$.

As seen in Fig.~\ref{fig:limit}, this value of $m_h$ is compatible with
$m_{1/2}
\lappeq$ 1400 GeV, as favoured by supersymmetric dark matter. It is
possible to strengthen the upper limit on $m_{1/2}$ if $m_h =
115$~GeV, $m_t = 175$ or 180
GeV and $\tan\beta$ is sufficiently large, but not if $m_t$ = 170 GeV. Of
course, these upper limits would evaporate with the Higgs signal. 

\begin{figure}
\vspace*{-0.25in}
\begin{minipage}{8in}
\epsfig{file=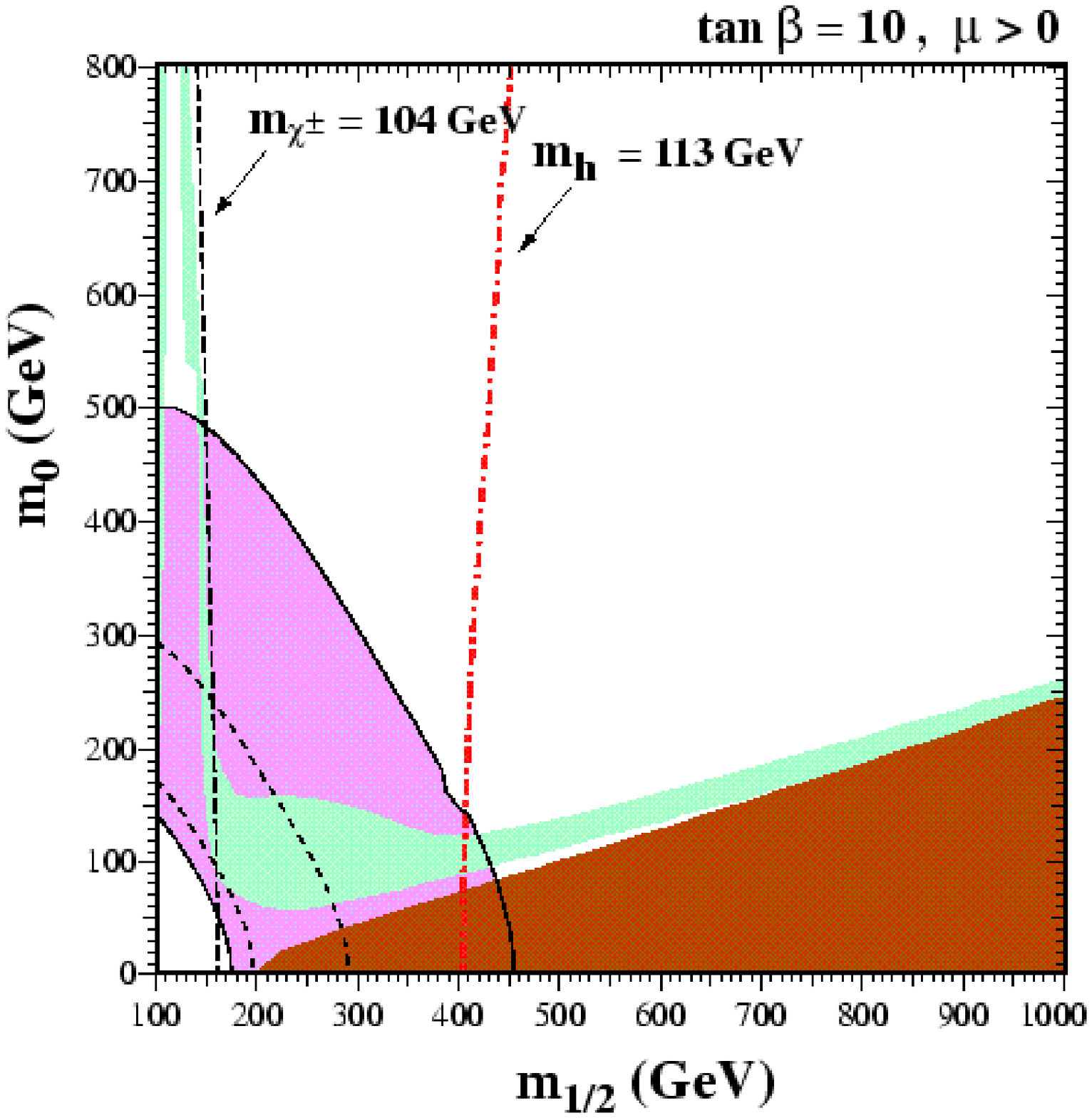,height=2.75in}
\hfill
\end{minipage}   
\begin{minipage}{8in}
\epsfig{file=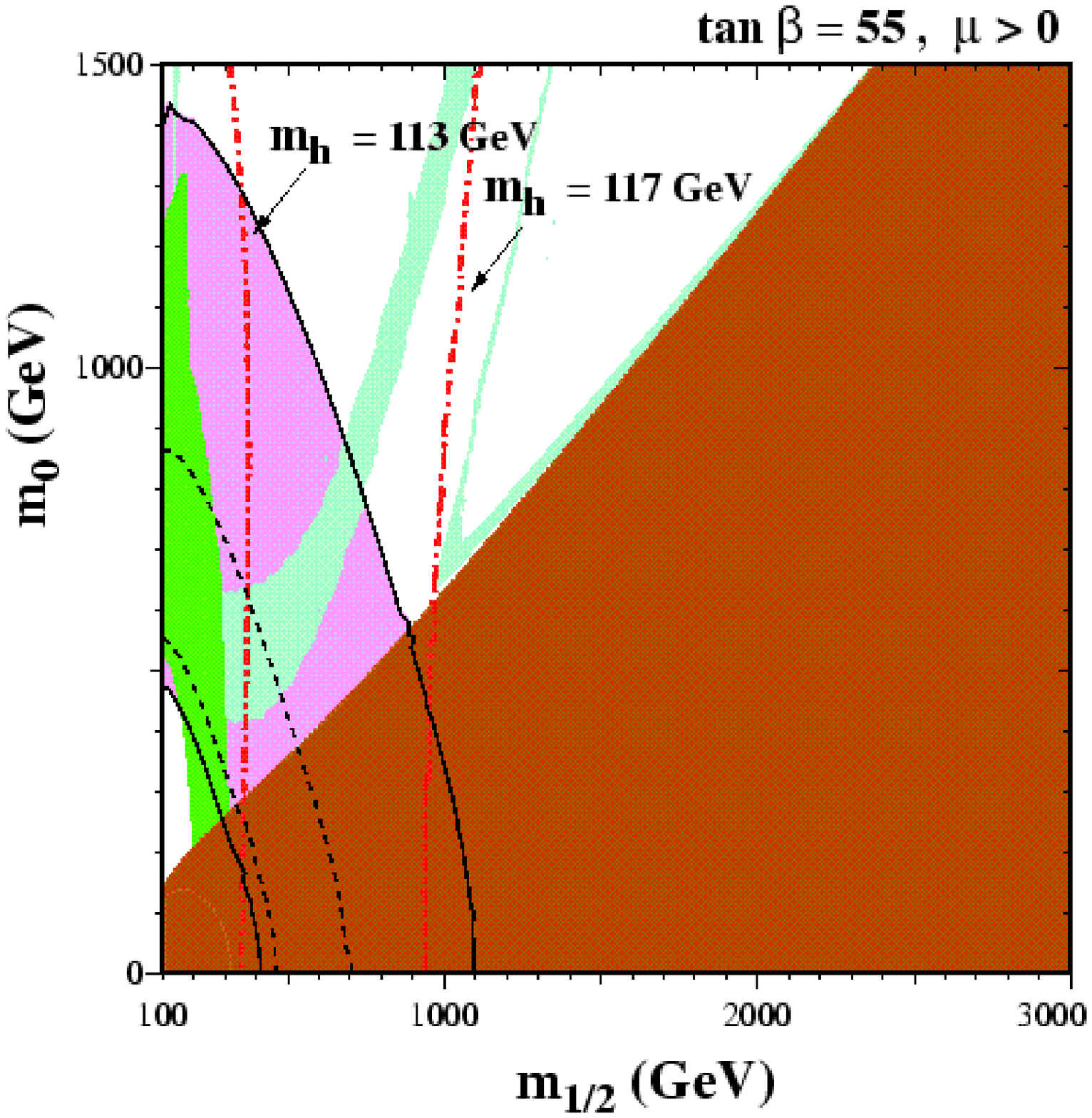,height=2.75in}
\hfill
\end{minipage}
\caption[.]{\label{fig:ENO}\it
Superpositions on the CMSSM $(m_{1/2}, m_0)$ planes for $\mu > 0$ and
$\tan
\beta = 10, 55$, see
Fig.~\ref{fig:EFGOSiplus}, of the constraint imposed by the E821
measurement
of
$a_\mu$. The regions allowed at the 2-$\sigma$ level are shaded (pink) and
bounded by solid black lines, with dashed lines indicating the 1-$\sigma$
ranges. 
} 
\end{figure}

Finally, I cannot resist adding as an anachronistic addendum a short
discussion of the possible evidence for physics beyond the Standard Model
provided by the recent BNL E821 measurement of the anomalous magnetic
moment of the muon. {\it Prima facie}, the discrepancy in this measurement
is evidence for new physics at the TeV scale, and supersymmetry could well
explain the E821 result, {\it if} $\mu > 0$. Fig.~\ref{fig:ENO} compares
the regions of the CMSSM parameter space allowed by cosmology, the LEP
sparticle exclusions, the Higgs `signal' and $b \to s \gamma$ decay with
that suggested by the E821 measurement. There is good consistency for
$\tan \beta \gappeq 10$ and moderate values of $m_{1/2}$ and $m_0$, well
within the reach of the LHC. The BNL E821 measurement does not change the
lower limit on $m_{1/2}$ and hence $m_\chi$ reported earlier, but it
would, if confirmed, strengthen the upper limit (\ref{upperchi}) to about
400~GeV. 

\section{How many Swallows?} 

We have seen in this talk many indirect indications that the summer of
supersymmetry may be on its way: the hierarchy problem, the unification of
gauge couplings, a (possible) light Higgs boson, cosmological dark matter,
and most recently the anomalous magnetic moment of the muon. These all
encourage us theorists who have been hibernating through 30 years of
supersymmetric winter. However, confirmation of our speculations can only
come from the direct detection of supersymmetric particles, either in an
accelerator experiment or as part of the astrophysical dark matter. If the
non-accelerator experiments do not find it first, I am optimistic that the
LHC will find supersymmetry, assuming that our phenomenological
motivations for expecting it at the TeV scale have any validity. If all
goes well, the LHC should be able to confirm or refute our speculations by
the end of the decade. Compared with the 30 years we have already waited,
perhaps that should not seem so long to wait.

\begin{flushleft}
{\bf Acknowledgements}\\
It is a pleasure to thank my many collaborators on the subjects discussed
here, most particularly Keith~Olive.
\end{flushleft}


\begin{thebibliography}{99}

\bibitem{hierarchy}
L.~Maiani, {\it Proceedings of the 1979 Gif-sur-Yvette Summer School On Particle
Physics}, 1;
G.~'t Hooft, in {\it Recent Developments In Gauge Theories, Proceedings 
of the Nato Advanced Study
Institute, Cargese, 1979}, eds. G.~'t Hooft {\it et al.}, (Plenum Press,
NY, 1980); E.~Witten,
Phys.\ Lett.\  {\bf B105} (1981) 267.

\bibitem{susyGUT}
J.~Ellis, S.~Kelley and D.~V.~Nanopoulos,
Phys.\ Lett.\  {\bf B249} (1990) 441 and
Phys.\ Lett.\  {\bf B260} (1991) 131;
U.~Amaldi, W.~de Boer and H.~Furstenau,
Phys.\ Lett.\  {\bf B260} (1991) 447;
C.~Giunti, C.~W.~Kim and U.~W.~Lee,
Mod.\ Phys.\ Lett.\  {\bf A6} (1991) 1745;
P.~Langacker and M.~Luo,
Phys.\ Rev.\  {\bf D44} (1991) 817.

\bibitem{lightH}
LEP Electroweak Working Group, \\
{\tt http://lepewwg.web.cern.ch/LEPEWWG/\\
stanmod/}.

\bibitem{EFZ}
J.  Ellis, G.  Ridolfi and F.  Zwirner, Phys.\ Lett.\
 {\bf B257} (1991) 83; M.S.  Berger, Phys.\ Rev.\ {\bf D41} (1990) 225;
 Y.  Okada, M.  Yamaguchi and T. Yanagida, Prog.\ Theor.\ Phys.\ {\bf
 85} (1991) 1; Phys.\ Lett.\ {\bf B262} (1991) 54; H.E.  Haber and R.
 Hempfling, Phys.\ Rev.\ Lett.\ {\bf 66} (1991) 1815.

\bibitem{EGNO}
J.~Ellis, G.~Ganis, D.~V.~Nanopoulos and K.~A.~Olive,
Phys.\ Lett.\ {\bf B502} (2001) 171.

\bibitem{BNL}
H.~N.~Brown {\it et al.}, BNL E821 Collaboration,
Phys.\ Rev.\ Lett.\ {\bf 86} (2001) 2227.

\bibitem{ENO}
J.~Ellis, D.~V.~Nanopoulos and K.~A.~Olive,
hep-ph/0102331.
For other supersymmetric analyses of the E821 result, see:
L.~Everett, G.~L.~Kane, S.~Rigolin and L.~Wang, hep-ph/0102145;
J.~L.~Feng and K.~T.~Matchev,
hep-ph/0102146;  
E.~A.~Baltz and
P.~Gondolo, hep-ph/0102147;
U.~Chattopadhyay and P.~Nath, hep-ph/0102157;
S.~Komine, T.~Moroi and M.~Yamaguchi, hep-ph/0102204;
R.~Arnowitt, B.~Dutta, B.~Hu and Y.~Santoso, hep-ph/0102344;
T.~Kobayashi and H.~Terao, hep-ph/0103028;
K.~Choi, K.~Hwang, S.~K.~Kang and K.~Y.~Lee,
hep-ph/0103048;
S.~P.~Martin and J.~D.~Wells,
hep-ph/0103067;
S.~Komine, T.~Moroi and M.~Yamaguchi, hep-ph/0103182;
K.~Cheung, C.~H.~Chou and O.~C.~W.~Kong, hep-ph/0103183;
S.~Baek, P.~Ko and H.~S.~Lee, hep-ph/0103218.

\bibitem{BBN}
K.~A.~Olive, G.~Steigman and T.~P.~Walker,
Phys.\ Rept.\  {\bf 333-334} (2000) 389.

\bibitem{CMBR}
A.~E.~Lange {\it et al.},
astro-ph/0005004.

\bibitem{nu}
J.~Ellis, summary talk at the {\it XIXth International Conference on
Neutrino
Physics and Astrophysics}, Sudbury, Canada, June 2000:
hep-ph/0008334 and references therein.

\bibitem{SF}
For a review, see:
E.~Gawiser and J.~Silk,
Science{\bf 280} (1998) 1405.

\bibitem{hiz}
See, for example, N.~Bahcall, J.~P.~Ostriker, S.~Perlmutter and
P.~J.~Steinhardt,
Science {\bf 284} (1999) 1481.

\bibitem{ERS}
J.~Ellis, D.~V.~Nanopoulos, L.~Roszkowski and D.~N.~Schramm,
Phys.\ Lett.\ {\bf B245} (1990) 251.

\bibitem{sarches}
For a review, see:
L.~Bergstrom,
Rept.\ Prog.\ Phys.\ {\bf 63} (2000) 793.

\bibitem{Fayet}
See, in particular, P. Fayet, Phys.\ Lett.\ {\bf 69B} (1977) 489. 
For a review of his early work, see: P.~Fayet,
in {\it Proceedings of the Europhysics Study Conf. on Unification of
Fundamental Interactions}, Erice, Italy, Mar 17-24, 1980.

\bibitem{Rviol}
See, for example:
H.~Dreiner,
hep-ph/9707435.

\bibitem{CDEO}
B.~A.~Campbell, S.~Davidson, J.~Ellis and K.~A.~Olive,
Phys.\ Lett.\ B {\bf 256} (1991) 457;
W.~Fischler, G.~F.~Giudice, R.~G.~Leigh and S.~Paban,
Phys.\ Lett.\ B {\bf 258} (1991) 45.

\bibitem{FF}
G.~R.~Farrar and P.~Fayet, Phys.\ Lett.\ {\bf 76B} (1978) 575 and {\bf
79B} (1978) 442.

\bibitem{Goldberg}
H. Goldberg, Phys. Rev. Lett. {\bf 50} (1983) 1419.

\bibitem{EHNS}
J.~Ellis, J.~S.~Hagelin, D.~V.~Nanopoulos and M.~Srednicki,
Phys.\ Lett.\ {\bf B127} (1983) 233.

\bibitem{Krauss}
L.~M.~Krauss,
Nucl.\ Phys.\ {\bf B227} (1983) 556.

\bibitem{EHNOS}
J. Ellis, J.S. Hagelin, D.V. Nanopoulos, K.A. Olive
and M. Srednicki, Nucl. Phys. {\bf B238} (1984) 453.

\bibitem{nocharged}
See, for example, J.~L.~Basdevant, R.~Mochkovitch, J.~Rich, M.~Spiro and
A.~Vidal-Madjar,
Phys.\ Lett.\  {\bf B234} (1990) 395.

\bibitem{EFGO}
J.~Ellis, T.~Falk, G.~Ganis and K.~A.~Olive,
hep-ph/0004169.
 
\bibitem{EFGOSi}
J.~Ellis, T.~Falk, G.~Ganis and K.~A.~Olive,
Phys.\ Rev.\ {\bf D62} (2000) 075010.

\bibitem{chargino}
LEP2 SUSY Working Group, \\
{\tt http://lepsusy.web.cern.ch/lepsusy/\\
www/inos\_moriond01/charginos\_pub.html}.

\bibitem{selectron}
LEP2 SUSY Working Group, \\
{\tt http://alephwww.cern.ch/${\tilde{~}}$ganis/SUSYWG/\\
SLEP/sleptons\_2k01.html}.

\bibitem{Higgs}
LEP Higgs Working Group, \\
{\tt  http://lephiggs.web.cern.ch/LEPHIGGS/\\
papers/mssm\_2001\_march/index.html}.

\bibitem{stop}
LEP2 SUSY Working Group, \\
{\tt http://lepsusy.web.cern.ch/lepsusy/www/\\
squarks\_moriond01/squarks\_pub.html}.

\bibitem{bsgamma}
CLEO Collaboration,
M.S. Alam et al., Phys.Rev.Lett. {\bf 74} (1995) 2885;
S.~Ahmed et al., {\tt CLEO CONF 99-10};
ALEPH Collaboration, R. Barate {\it et al.}, Phys.
Lett. {\bf B429} (1998) 169;
BELLE Collaboration, K.~Abe {\it et al.}, hep-ex/0103042.

\bibitem{NLObsg}
G.~Degrassi, P.~Gambino and G.~F.~Giudice,
hep-ph/0009337;
M.~Carena, D.~Garcia, U.~Nierste and C.~E.~Wagner,
hep-ph/0010003.

\bibitem{CCB}
J.~A.~Casas, A.~Lleyda and C.~Munoz,
Nucl.\ Phys.\  {\bf B471} (1996) 3;
H. Baer, M. Brhlik and D. Castano, Phys. Rev. {\bf
D54} {6944} {1996};
S.~Abel and T.~Falk, Phys.\ Lett.\  {\bf B444} (1998) 427.

\bibitem{EFOSi}
J.~Ellis, T.~Falk and  K.~A.~Olive, Phys.\ Lett.\  {\bf B444}, 3;
J.~Ellis, T.~Falk, K.~A.~Olive and M.~Srednicki,
Astropart.\ Phys.\  {\bf 13} (2000) 181.

\bibitem{othercoann}
For another treatment of $\chi - {\tilde \tau}$ coannihilation, see:
M.~E.~G\'omez,
G.~Lazarides and C.~Pallis,
Phys.\ Rev.\ {\bf D61}, 123512 (2000)
and
Phys.\ Lett.\ {\bf B487}, 313 (2000).   
For $\chi - \chi^\prime - \chi^\pm$ coannihilation, see:
S.~Mizuta and M.~Yamaguchi,
Phys.\ Lett.\ {\bf B298} (1993) 120;
J.~Edsjo and P.~Gondolo,
Phys.\ Rev.\ {\bf D56} (1997) 1879.
For $\chi - {\tilde t}$ coannihilation, see:
C.~Boehm, A.~Djouadi and M.~Drees,
Phys.\ Rev.\ {\bf D62} (2000) 035012.
Neither of the latter processes is important in the CMSSM, though they may
be relevant in the more general MSSM.

\bibitem{LHC}
ATLAS Collaboration, Detector and Physics Performance Technical Design
Report, \\
{\tt http://atlasinfo.cern.ch/Atlas/GROUPS/ PHYSICS/TDR/access.html};\\
CMS Collaboration, Technical Proposal,\\
{\tt http://cmsinfo.cern.ch/TP/TP.html}.

\bibitem{Tevatron}
See S.~Abel {\it et al.}, Tevatron SUGRA Working Group Collaboration,
hep-ph/0003154, and references therein.

\bibitem{halo}
J.~Silk and M.~Srednicki,
Phys.\ Rev.\ Lett.\  {\bf 53} (1984) 624.

\bibitem{pbars}
L.~Bergstrom, J.~Edsjo and P.~Ullio,
astro-ph/9902012.

\bibitem{clumpy}
L.~Bergstrom, J.~Edsjo, P.~Gondolo and P.~Ullio,
Phys.\ Rev.\  {\bf D59} (1999) 043506.

\bibitem{Sun}
J.~Silk, K.~Olive and M.~Srednicki,
Phys.\ Rev.\ Lett.\  {\bf 55} (1985) 257.

\bibitem{Gondolo}
P. Gondolo, talk at the {\it XIXth International Conference on Neutrino
Physics and Astrophysics}, Sudbury, Canada, June 2000:\\
{\tt http://nu2000.sno.laurentian.ca/}.

\bibitem{GW}
M.~W.~Goodman and E.~Witten,
Phys.\ Rev.\  {\bf D31} (1985) 3059.

\bibitem{DAMA}
R.~Bernabei {\it et al.}, DAMA Collaboration,
Phys.\ Lett.\  {\bf B480} (2000) 23.

\bibitem{EFO1}
J.~Ellis, A.~Ferstl and K.~A.~Olive,
Phys.\ Lett.\  {\bf B481} (2000) 304.

\bibitem{CDMS}
R.~Abusaidi {\it et al.}, CDMS Collaboration,
Nucl.\ Instrum.\ Meth.\  {\bf A444} (2000) 345.

\bibitem{EFO2}
J.~Ellis, A.~Ferstl and K.~A.~Olive,
Phys.\ Rev.\ {\bf D63} (2001) 065016.

\bibitem{Bottino}
For other calculations and more references, see: 
R.~Arnowitt, B.~Dutta and Y.~Santoso,
hep-ph/0101020;  
H.~Baer and M.~Brhlik,
Phys.\ Rev.\ D {\bf 57} (1998) 567;
A.~Bottino, F.~Donato, N.~Fornengo, S.~Scopel, hep-ph/0010203;
A.~Corsetti and P.~Nath,
hep-ph/0011313;
J.~L.~Feng, K.~T.~Matchev and F.~Wilczek,
Phys.\ Rev.\ D {\bf 63} (2001) 045024.

\bibitem{EFO3}
J.~Ellis, A.~Ferstl and K.~A.~Olive, in preparation.

\bibitem{LEPHiggs}
ALEPH collaboration, R.~Barate {\it et al.}, Phys.\ Lett.\ {\bf B495}
(2000) 1;\\
L3 collaboration, M.~Acciarri {\it et al.}, Phys.\ Lett.\ {\bf B495}
(2000) 18;\\
DELPHI collaboration, P. Abreu {\it et al.},
Phys.\ Lett.\ B {\bf 499} (2001) 23;\\
OPAL collaboration, G. Abbiendi {\it et al.}, Phys.\ Lett.\ {\bf B499}
38.

\bibitem{PIK}
P. Igo-Kemenes, LEP Higgs working group,\\
{\tt http://lephiggs.web.cern.ch/LEPHIGGS/\\
talks/index.html}.

\bibitem{TC}
E.~Farhi and L.~Susskind,
Phys.\ Rept.\  {\bf 74} (1981) 277.

\bibitem{DE}
S.~Dimopoulos and J.~Ellis,
Nucl.\ Phys.\  {\bf B182} (1982) 505.

\bibitem{ER}
J.~Ellis and D.~Ross,
hep-ph/0012067.

\end{thebibliography}
\end{document}